\documentclass[final
  ]
  {aipproc}

\layoutstyle{6x9}

\begin{document}

\vspace*{-3em}
\hfill JLAB-THY-10-1165

\title{Hypernuclei in the quark-meson coupling model}


\author{K. Tsushima}{
  address={Thomas Jefferson Lab., 12000 Jefferson Ave., Newport News, VA 2
3606, USA}
}

\author{P.~A.~M. Guichon}{
  address={SPhN-DAPNIA, CEA Saclay, F91191 Gif sur Yvette, France}
}


\begin{abstract}
We present results of hypernuclei calculated in the latest quark-meson
coupling (QMC) model, where the effect of the mean scalar field
in-medium on the one-gluon exchange hyperfine interaction,
is also included self-consistently.
The extra repulsion associated with this increased hyperfine interaction
in-medium completely changes the predictions for $\Sigma$ hypernuclei. Whereas in
the earlier version of QMC they were bound by an amount similar to $\Lambda$ hypernuclei,
they are unbound in the latest version of QMC, in qualitative agreement with the experimental
absence of such states.
\end{abstract}

\maketitle


\section{Introduction}

The study of $\Lambda$ hypernuclei has provided us with important information
on the properties of $\Lambda$ in a nuclear medium and the effective $\Lambda$-N
interaction~\cite{Hashimoto:2006aw}. However, the situation for $\Sigma$ and $\Xi$
hypernuclei is quite different.
The special case of $^4_{\Sigma}$He aside, there is no experimental
evidence for any $\Sigma$ hypernuclei~\cite{Saha:2004ha},
despite extensive searches.
It seems likely that the $\Sigma$-nucleus interaction is somewhat repulsive and that
there are no bound $\Sigma$ hypernuclei beyond A=4. In the case of the $\Xi$,
the experimental situation is very challenging, but we eagerly await studies of
$\Xi$ hypernuclei with new facilities at J-PARC and GSI-FAIR.

To understand further the properties of hypernuclei, we have used the latest version of
the quark-meson coupling (QMC) model~\cite{RikovskaStone:2006ta}, which will be referred to as
QMC-III, and computed the single-particle energies~\cite{qmcGTT}.
(The earliest version of QMC will be referred to as QMC-I, where QMC-II~\cite{qmc2} also exists.)
The major improvement in the QMC-III model is the inclusion of the effect
of the medium on the color-hyperfine interaction. This has the effect of increasing the splitting
between the $\Lambda$ and $\Sigma$ masses as the density rises. This is the prime
reason why our results yield no middle and heavy mass $\Sigma$ hypernuclei~\cite{qmcGTT}.

The QMC model was created to provide insight into the structure of nuclear matter,
starting at the quark level~\cite{Guichon:1987jp,qmcPPNPreview}.
Nucleon internal structure was modeled by the MIT bag,
while the binding was described by the self-consistent couplings of the
confined light quarks ($u,d$) (not $s$ nor heavier quarks!) to the scalar-$\sigma$ and vector-$\omega$
meson fields generated by the confined light quarks in the other ``nucleons''.
The self-consistent response of the bound light quarks to the mean $\sigma$ field
leads to a novel saturation mechanism for nuclear matter, with the enhancement of
the lower components of the valence Dirac quark wave functions.
The direct interaction between the light quarks and the scalar
$\sigma$ field is the key of the model, which induces the {\it scalar polarizability}
at the nucleon level, and generates the nonlinear scalar potential (effective nucleon mass), or
the density ($\sigma$-field) dependent $\sigma$-nucleon coupling.
The model has opened tremendous opportunities for the studies of finite nuclei
and hadron properties in a nuclear medium (in nuclei), based on the quark degrees
of freedom~\cite{qmcPPNPreview}.

\section{Hyperons in nuclear matter}

\begin{figure}[htb]
\hspace*{1.5em}
\includegraphics[angle=-90,width=.5\textwidth]{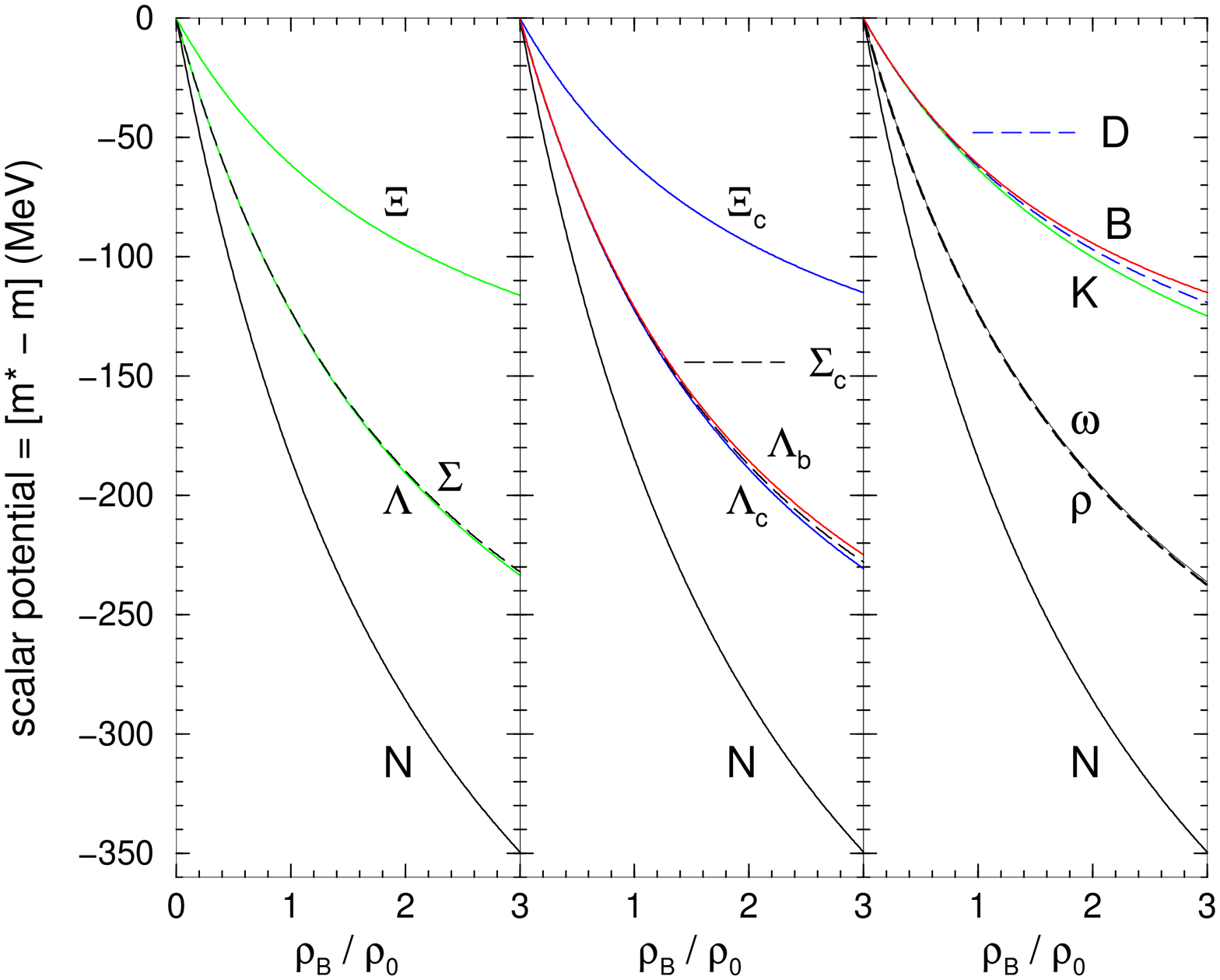}
\includegraphics[angle=-90,width=.55\textwidth]{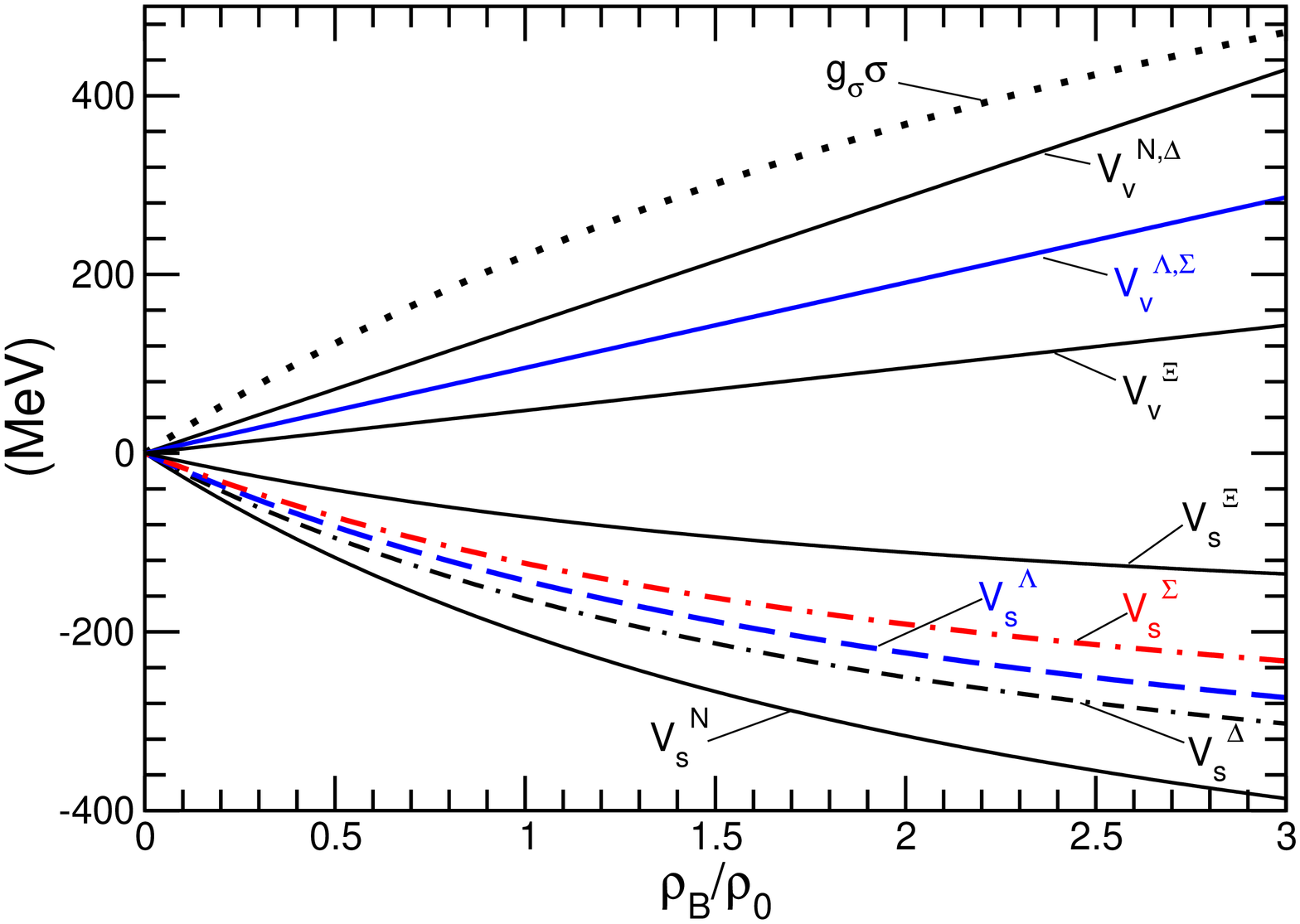}
\caption{Scalar potentials in QMC-I (left panel)~\cite{QMCbcmatter}, and scalar ($V_S$)
and vector ($V_V$) potentials in QMC-III (right panel)~\cite{qmcGTT},
in symmetric nuclear matter. The vector potentials are the same for both QMC-I and QMC-III,
proportional to the light quark number in a hadron, and liner as a function of baryon density.}
\label{spot}
\end{figure}

Since the coupling constants of the light quarks ($u,d$) and
$\sigma$, $\omega$, and $\rho$ fields are the same for all the light quarks in any hadrons
in QMC, the model can treat the interactions in a systematic, unified manner.
In particular, the scalar potentials (in-medium mass minus free mass)
for hadrons in QMC-I have turned out to be proportional to the light quark number
in a hadron --- the light quark number counting rule~\cite{QMCbcmatter}.
This is shown in the left panel in Fig.~\ref{spot}.

As one can see from the left panel in Fig.~\ref{spot}, the attractive scalar potentials felt
by the $\Lambda$ and $\Sigma$ are nearly the same.
Since the repulsive vector potential is proportional exactly to the light quark number in QMC,
the total, nonrelativistic potentials felt by the $\Lambda$ and $\Sigma$ are
very similar.
Thus, as in usual SU(3)-based relativistic mean field models, this
naturally led to predict the existence of bound $\Sigma$ hypernuclei in QMC-I~\cite{qmchyp},
with the similar amount with that of the $\Lambda$, despite of some deviations
due to the $\Lambda-\Sigma$ channel coupling and phenomenologically introduced
Pauli blocking effect at the quark level~\cite{qmchyp}.

However, this difficulty, which contradicts to the experimental observations,
is resolved in QMC-III~\cite{qmcGTT}. It is the self-consistent inclusion of the color-hyperfine
interaction in a nuclear medium that resolves this difficulty.
(Based on the quark and gluon dynamics!)
By this color-hyperfine interaction in the nuclear medium,
the scalar potential for the $\Lambda$ gets more attraction, while that for the $\Sigma$ gets
less attraction. (Similarly,the scalar potential for the $\Delta$ becomes less attractive than that
for the nucleon.) The scalar ($V_S$) and vector ($V_V$) potentials calculated
in QMC-III in symmetric nuclear matter are shown in the right panel in Fig.~\ref{spot}.

Explicit expressions for the effective masses (in-medium masses) in QMC-III are,
\begin{eqnarray}
M_{N}(\sigma) & = & M_{N}
- g_{\sigma}\sigma\nonumber \\
 &  & +\left[0.002143+0.10562R_{N}^{free}
-0.01791\left(R_{N}^{free}\right)^{2}\right]
\left(g_{\sigma}\sigma\right)^{2},
\label{eq:mN}\\
M_{\Delta}(\sigma) & = & M_{\Delta}
- \left[ 0.9957 - 0.22737R_{N}^{free} +0.01\left(R_{N}^{free}\right)^{2}\right] g_{\sigma}\sigma \nonumber\\
 &  & +\left[0.0022 + 0.1235R_{N}^{free}
- 0.0415\left(R_{N}^{free}\right)^{2}\right]
\left(g_{\sigma}\sigma\right)^{2},
\label{eq:mD}\\
M_{\Lambda}(\sigma) & = & M_{\Lambda}-\left[0.6672+0.04638R_{N}^{free}-
0.0022\left(R_{N}^{free}\right)^{2}\right]
g_{\sigma}\sigma\nonumber \\
 &  & +\left[0.00146+0.0691R_{N}^{free}-0.00862
\left(R_{N}^{free}\right)^{2}\right]
\left(g_{\sigma}\sigma\right)^{2},
\label{eq:mL}\\
M_{\Sigma}(\sigma) & = & M_{\Sigma}-\left[0.6653-0.08244R_{N}^{free}+
0.00193\left(R_{N}^{free}\right)^{2}\right]
g_{\sigma}\sigma\nonumber \\
 &  & +\left[0.00064+0.07869R_{N}^{free}-0.0179
\left(R_{N}^{free}\right)^{2}\right]
\left(g_{\sigma}\sigma\right)^{2},
\label{eq:mS}\\
M_{\Xi}(\sigma) & = & M_{\Xi}-\left[0.3331+0.00985R_{N}^{free}-
0.00287\left(R_{N}^{free}\right)^{2}\right]
g_{\sigma}\sigma\nonumber \\
 &  & +\left[-0.00032+0.0388R_{N}^{free}-
0.0054\left(R_{N}^{free}\right)^{2}\right]
\left(g_{\sigma}\sigma\right)^{2}\, ,
\label{eq:mX}
\end{eqnarray}
where, the bag radius in free sapce, $R_N^{free}$, has been taken 0.8 fm for numerical calculations,
but the results are quite insensitive
(c.f. Fig.~1 of Ref.~\cite{RikovskaStone:2006ta}) to this parameter.

\section{Hypernuclei}

In this section we present the results for hypernuclei calculated in QMC-III.
Details are given in Ref.~\cite{qmcGTT}.
To calculate the hyperon levels, we use a relativistic shell model, and
generate the shell model core using the Hartree approximation. The free space meson
nucleon coupling constants are,
$g_\sigma^2 = 8.79 m_\sigma^2$, $g_\omega^2 = 4.49 m_\omega^2$
and $g_\rho^2 = 3.86 m_\rho^2$, with $m_\sigma$ = 700 MeV, $m_\omega$ =
770 MeV and $m_\rho = 780$ MeV~\cite{RikovskaStone:2006ta}.
Once we have the shell model core wave functions,
we use the more sophisticated Hartree-Fock couplings for the hyperon.
In a previous study of high central density
neutron stars~\cite{RikovskaStone:2006ta}, where the hyperon population
is large enough that their exchange terms matter, we found that the
Hartree-Fock couplings, $g_\sigma^2 = 11.33 m_\sigma^2$,
$g_\omega^2 = 7.27 m_\omega^2$ and $g_\rho^2 = 4.56 m_\rho^2$,
gave a satisfactory phenomenology. So, for the hyperons we use
these couplings. (See also Eqs.~(\ref{eq:mL}) -~(\ref{eq:mX})).

Before discussing the results in detail, we first note the remarkable agreement
between the calculated ($-26.9$ MeV in $^{209}_{\Lambda}$Pb) and the experimental
($-26.3 \pm 0.8$ MeV in $^{208}_{\Lambda}$Pb) binding energy of the $\Lambda$ in the $1s_{1/2}$ level.
In our earlier work the $\Lambda$ was
overbound by 12 MeV and we needed to add a
phenomenological correction which we attributed to
the Pauli effect at the quark level.
This correction is not needed when we use Hartree-Fock, rather
than Hartree, coupling constants.

Already at this stage the binding of the $\Sigma^0$ in the $1s_{1/2}$
level of $^{209}_{\Sigma^0}$Pb is just a few MeV -- a major improvement over the earlier
QMC-I results. However, there is an additional piece of physics which really
should be included and which goes beyond the
naive description of the intermediate range attraction in terms of
$\sigma$ exchange. In particular, the energy released in the
two-pion exchange process, N $\Sigma \to$ N $\Lambda \to$ N $\Sigma$,
because of the $\Sigma$--$\Lambda$ mass difference,
reduces the intermediate range attraction felt by the $\Sigma$ hyperon.
In Ref.~\cite{qmchyp} this was modeled
by introducing an additional
vector repulsion for a $\Sigma$ hyperon. Following the same
procedure, we replace $g_\omega^{\Sigma} \omega(r)$
by $g_\omega^{\Sigma} \omega(r)
+ \lambda_\Sigma \rho_B(r)$,
with $\lambda_\Sigma = 50.3$ MeV-fm$^3$, as determined
in Ref.~\cite{qmchyp} by the comparison with
the more microscopic model of the J\"ulich group~\cite{Reuber:1993ip}.

Our results are presented in Tables~\ref{spe1} and~\ref{spe2}. The overall
agreement with the experimental energy levels of $\Lambda$ hypernuclei
across the periodic table is quite good. The discrepancies
which remain may well be
resolved by small effective hyperon-nucleon interactions which go beyond
the simple, single-particle shell model. Once again, we stress the very
small spin-orbit force experienced by the $\Lambda$, which is a natural
property of the QMC model~\cite{qmchyp}.

\begin{table}[htbp]
{\begin{tabular}{c|ccc|ccc|cc}
\hline
&$^{16}_\Lambda$O(Exp.) &$^{17}_\Lambda$O &$^{17}_{\Xi^0}$O
&$^{40}_\Lambda$Ca(Exp.)&$^{41}_\Lambda$Ca &$^{41}_{\Xi^0}$Ca
&$^{49}_\Lambda$Ca     &$^{49}_{\Xi^0}$Ca\\
\hline
$1s_{1/2}$&-12.42 &-16.2 &-5.3 &-18.7 &-20
.6 &-5.5 &-21.9 &-9.4 \\
&$\pm0.05$& & &$\pm 1.1$& & & & \\
&$\pm0.36$& & & & & & & \\
$1p_{3/2}$&           & -6.4 &---  &            &-13.9 &-1.6 &-15.4 &-5.3 \\
$1p_{1/2}$& -1.85 & -6.4 &---  & &-13.9 &-1.9 &-15.4 &-5.6\\
&$\pm 0.06$& & & & & & & \\
&$\pm 0.36$& & & & & & & \\
\hline
\end{tabular}}
\caption{Single-particle energies (in MeV)
for $^{17}_Y$O, $^{41}_Y$Ca and $^{49}_Y$Ca hypernuclei ($Y=\Lambda,\Xi^0$).
Neither the $\Sigma^0$ nor the $\Sigma^+$ is bound in strong interaction.
The experimental data are taken from Ref.~\protect\cite{Hashimoto:2006aw} (Table 11) for
$^{16}_\Lambda$O and from Ref.~\protect\cite{Pile:1991cf} for $^{40}$Ca.}
\label{spe1}
\end{table}
\begin{table}[htbp]
{\begin{tabular}{c|ccc|ccc}
\hline
&$^{89}_\Lambda$Yb(Exp.)  &$^{91}_\Lambda$Zr  &$^{91}_{\Xi^0}$Zr
&$^{208}_\Lambda$Pb(Exp.) &$^{209}_\Lambda$Pb &$^{209}_{\Xi^0}$Pb \\
\hline
$1s_{1/2}$&-23.1 $\pm 0.5$       &-24.0 &-9.9 &-26.3 $\pm 0.8$&-26.9 &-15.0\\
$1p_{3/2}$&            &-19.4 &-7.0 &            &-24.0 &-12.6 \\
$1p_{1/2}$&-16.5 $\pm 4.1$ ($1p$)&-19.4 &-7.2 &-21.9 $\pm 0.6$ ($1p$)&-24.0&-12.7 \\
$1d_{5/2}$&            &-13.4 &-3.1 &---         &-20.1 & -9.6 \\
$2s_{1/2}$&            & -9.1 &---  &---         &-17.1 & -8.2 \\
$1d_{3/2}$&-9.1 $\pm 1.3$  ($1d$)&-13.4 &-3.4 &-16.8 $\pm 0.7$ ($1d$)&-20.1& -9.8 \\
$1f_{7/2}$&            & -6.5 &---  &---         &-15.4 & -6.2 \\
$2p_{3/2}$&            & -1.7 &---  &---         &-11.4 & -4.2 \\
$1f_{5/2}$&-2.3 $\pm1.2$  ($1f$)& -6.4 &---  &-11.7 $\pm 0.6$ ($1f$)&-15.4 &-6.5 \\
$2p_{1/2}$&            & -1.6 &---  &---         &-11.4 & -4.3 \\
$1g_{9/2}$&            &---   &---  &---         &-10.1 & -2.3 \\
$1g_{7/2}$&            &---   &---  & -6.6 $\pm 0.6$ ($1g$)&-10.1 & -2.7 \\
\hline
\end{tabular}}
\caption{Same as table~\ref{spe1} but
for $^{91}_Y$Zr and $^{209}_Y$Pb hypernuclei. The experimental data are
taken from Ref.~\protect\cite{Hashimoto:2006aw} (Table 13).}
\label{spe2}
\end{table}

\begin{figure}[htb]
  \includegraphics[height=.35\textheight,width=.5\textwidth]{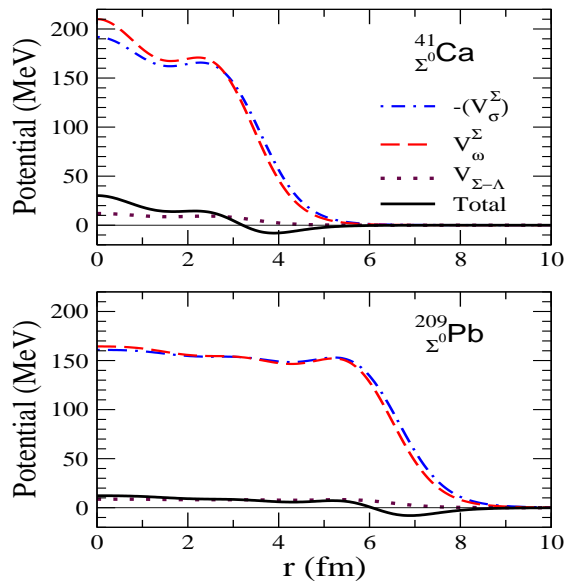}
  \caption{$\Sigma^0$ potentials in $^{41}_{\Sigma^0}$Ca and $^{209}_{\Sigma^0}$Pb.
  See also Refs.~\cite{qmcGTT,qmchyp} for detail.}
\label{Sig0pot}
\end{figure}

There are no entries for the $\Sigma$-hyperon because neither the $\Sigma^+$
nor the $\Sigma^0$ is bound to a finite nucleus in strong interaction.
This absence of bound $\Sigma$ hypernuclei constitutes a major advance over
QMC-I. We stress that this is a direct consequence of the enhancement of the hyperfine
interaction (that splits the masses of the $\Sigma$ and $\Lambda$ hyperons) by the
mean scalar field in-medium. It is especially interesting to examine the effective
non-relativistic potential felt by the $\Sigma^0$ in a finite nucleus.
For example, we show in Fig.~\ref{Sig0pot} the $\Sigma^0$ potentials
in $^{41}_{\Sigma^0}$Ca and $^{209}_{\Sigma^0}$Pb.
In $^{41}_{\Sigma^0}$Ca, the vector repulsion from
the $\omega$ wins in the center, with the
potential being as large as $+20$ to $+30$ MeV there,
while the scalar attraction wins in the surface with the
potential reaching approximately $-10$ MeV around 4fm.
For $^{209}_{\Sigma^0}$Pb, the trend is similar.

\vspace*{3em}

While the exact numerical values depend on the
mass taken for the $\sigma$ meson, we stress
the similarity to
the phenomenological form found by
Batty {\it et al.}~\cite{Batty:1994sw}. For a recent review
see~\cite{Friedman:2007qx}. It will clearly be very interesting to pursue the
application of the current theoretical formulation
to $\Sigma^-$-atoms.

We also note that this model supports the existence of a variety of
bound $\Xi$ hypernuclei. For the $\Xi^0$ the binding
of the 1s level varies from 5 MeV in $^{17}_{\Xi^0}$O to
15 MeV in $^{209}_{\Xi^0}$Pb. The experimental
search for such states at facilities such as
J-PARC and GSI-FAIR will be very important.

\section{Conclusion}

First, the inclusion of the effect of the medium on the one-gluon exchange
color magnetic hyperfine interaction between quarks within the quark-meson coupling model (QMC-III),
has led to some important advances. This is based on the quark and gluon dynamics, and
it would be non-trivial task for usual SU(3) symmetry and hadron based relativistic mean field
approaches to accommodate such effects leading to the absence of middle and
heavier mass $\Sigma$ hypernuclei.

Second, the agreement between the parameter free calculations and
the low-lying experimental energy levels for the $\Lambda$ hypernuclei is impressive,
especially between the calculated ($-26.9$ MeV in $^{209}_\Lambda$Pb)
and the experimental ($-26.3 \pm 0.8$ MeV in $^{208}_\Lambda$Pb)
single-particle energy of the $\Lambda$ in the $1s_{1/2}$ level.
However, for the d- and f-wave levels shown in Table~\ref{spe2}, there is a tendency for the
model to overbind by several MeV. Whether this is a consequence of the use of an extreme single
particle shell model for the core, the omission of residual $\Lambda-N$ interactions or an aspect
of the current implementation in QMC-III that requires improvement remains to be seen.

Third, a number of $\Xi$ hypernuclei are predicted to be bound, although not as deeply as in
the $\Lambda$ case.

Last, we emphasize again that the additional repulsion arising from the enhancement of the hyperfine
repulsion in the $\Sigma$-hyperon in-medium, together with the effect of the $\Sigma N -
\Lambda N$ channel coupling on the intermediate range scalar attraction, means that
no middle and heavy mass $\Sigma$ hypernuclei are predicted to be bound. This encouraging picture of finite
hypernuclei, suggests that the underlying model, which is fully relativistic and incorporates the
quark substructure of the baryons, is ideally suited for application to the properties of
dense matter and neutron stars.


\begin{theacknowledgments}
KT would like to acknowledge CSSM for hospitality,
and thank A.~W.~Thomas for many exciting collaborations, and celebrate his 60th birthday!
Notice: Authored by Jefferson Science Associates, LLC under U.S. DOE Contract No. DE-AC05-06OR23177.
The U.S. Government retains a non-exclusive, paid-up, irrevocable, world-wide license to publish
or reproduce this manuscript for U.S. Government purposes.
\end{theacknowledgments}


\bibliographystyle{aipproc}   

\end{document}